\documentclass[12pt]{article}
\usepackage{authblk}
\usepackage{graphicx}
\usepackage{amssymb}
\usepackage{amsmath}
\title{
FRW universe in the laboratory
 }
\author[1,2]{Neven Bili\'c\thanks{bilic@irb.hr}}
\affil[1]{Rudjer Bo\v skovi\'c Institute, 10002 Zagreb, Croatia}
\author[1]{Dijana Toli\'c\thanks{dijana.tolic@irb.hr}}
\affil[2]{Departamento de F\'isica,
Universidade Federal de Juiz de Fora, 36036-330,
Juiz de Fora, Minas Gerais, Brazil}

\date{\today}
\begin{document}
\maketitle
\begin{abstract}

 We consider an expanding relativistic fluid with spherical symmetry as a model 
 for an analog Friedmann-Robertson-Walker (FRW) spacetime. 
In the framework of relativistic acoustic geometry we demonstrate how to mimic 
 an arbitrary FRW spacetime with positive, zero  or negative spatial curvature.
 In the Lagrangian description we show  that a particular
 FRW spacetime is obtained by choosing the appropriate potential.
 We discuss several examples and in particular
 the analog  de Sitter spacetime in the coordinate representation
 with positive and negative spatial curvature.

\end{abstract}

%\pacs{03.75.Kk, 04.40.Nr, 47.10.-g, 98.80.Jk}
%
%\keywords{analog gravity, cosmology, fluid dynamics, Bose-Einstein condensate}

%\maketitle

\section{Introduction}
%One of the essential ingridients of General Relativity is the  geometry of spacetime.
The possibility that a curved (pseudo-Riemannian) geometry of spacetime
can be mimicked by fluid dynamics in Minkowski spacetime has been recently exploited in various contexts
including 
emergent gravity \cite{babichev}, scalar theory of gravity \cite{novello}, 
acoustic geometry \cite{visser,bilic,kinoshita} 
to name but a few.

The basic idea is the emergence of an effective metric of the form
\begin{equation}
G_{\mu\nu} = a g_{\mu\nu}+bu_{\mu}u_{\nu},
\label{eq1a}
\end{equation}
which describes the effective geometry for acoustic perturbations propagating in 
a perfect fluid potential flow with  $u_\mu\propto \partial_\mu \phi$.
The background spacetime metric $g_{\mu\nu}$ is usually assumed flat and 
the coefficients $a$ and $b$ are 
related to the equation of state of the fluid and the adiabatic speed of sound. Equivalently, 
in a field-theoretical picture the fluid velocity $u_\mu$ is derived from the scalar field as 
$u_\mu= \partial_\mu \phi/\sqrt{X}$ and $a$ and $b$ are expressed in terms of the Lagrangian and its first 
and second derivatives with respect to the kinetic energy term $X=g^{\mu\nu}\phi_{,\mu}\phi_{,\nu}$.

In a slightly different context,
the metric of the form (\ref{eq1a}) has been used to show that a pseudo-Riemann 
spacetime with Lorentz signature may be derived from a Riemann metric with Euclidean signature 
\cite{barbero1,barbero2, mukohyama}. In that case, the vector $u_\mu$ represents the normalized
gradient of a hypothetical scalar field which governs the dynamics and the signature of the effective
spacetime.

The effective geometry of an expanding fluid  seems a promising route to model
an expanding spacetime, e.g., of Friedmann-Robertson-Walker (FRW) type.
However, an expanding fluid alone is not a sufficiently powerful tool for  
modeling  an expanding spacetime of general relativity. 
At best, the fluid flow can be manipulated
to provide a coordinate transformation 
to an arbitrary
curved coordinate frame, with the spacetime remaining flat. For example,
a boost-invariant Bjorken-type spherical expansion provides  a map
of the Minkowski 
to the Milne spacetime -- a homogeneous, isotropic, expanding universe with the cosmological scale 
proportional to $t$ and negative spatial curvature 
\cite{lampert}.

In this paper we propose a mechanism in which the  metric (\ref{eq1a})
provides a laboratory model for an analog FRW spacetime with arbitrary spatial curvature.
To model an arbitrary FRW spacetime we make use of the effective acoustic geometry
in which small perturbations propagate adiabatically.
We show quite generally that in addition to
an appropriately designed fluid expansion it is necessary to
manipulate the fluid equation of state to obtain the desired geometry.
By making use of a specific type of fluid expansion in Minkowski spacetime 
it is possible quite generally  to map  
the Minkowski metric  into  
spatially closed ($k=1$), open hyperbolic ($k=-1$), or open flat ($k=0$) FRW acoustic metrics.
The mappings to spatially flat and hyperbolic universes have been 
already studied in a few specific applications, such as
spatially flat analog cosmology in a non-relativistic Bose-Einstein (BE) condensate
\cite{barcelo} and an open FRW metric in a Bjorken-like spherical
 expansion of a relativistic BE system \cite{fagnocchi}
and of a  hadronic fluid \cite{bilic2,bilic3}. 
We extend these ideas to a more general case that includes all three types of
FRW spatial geometries.
 It turns out that an open FRW spacetime with zero or negative spatial
curvature can be modeled in a rather straightforward way 
using isentropic perfect fluids such as a Bose-Einstein condensate in the Thomas-Fermi
limit.
The modeling of a closed FRW
spacetime with positive spatial curvature
requires additional assumptions.

We divide the remainder of the paper into five sections and an appendix
starting with section  \ref{spacetimes}, in which  
we give a field-theoretical description of our model.
In the following section, section \ref{bose}, we study the relativistic BE
condensate in the context of analog cosmology. Section \ref{hydrodynamic} 
reproduces the results of 
section \ref{spacetimes} in terms of the conventional relativistic hydrodynamics.
In section \ref{analog} we study the analog cosmological horizons
and we present the spacetime diagrams for the analog de Sitter universe.
We summarize our results and give  
a brief outlook in section \ref{discussion}. 
Finally, in appendix \ref{coordinate}
we describe a general 
transformation from
  Minkowski 
to spatially curved conformal 
coordinates.   
 
\section{Analog spacetimes}
\label{spacetimes}
We start with an expanding perfect fluid in Minkowski spacetime 
 in spherical coordinates $(T,R,\vartheta,\varphi)$, 
\begin{equation}
ds^2=dT^2-dR^2-R^2d\Omega^2,
\label{eq001a}
\end{equation}
and we demand that the effective acoustic metric in comoving coordinates 
$(t,r,\vartheta,\varphi)$ takes the conformal FRW form with line element
\begin{equation}
ds^2=\alpha(t)^2\left(c_s^2dt^2-dr^2-\frac{\sin^2(\sqrt{k}r)}{k}d\Omega^2\right).
%\label{eq3025}
\end{equation} 
Here $c_s$ is the (generally time-dependent) speed of sound
and  the curvature $k$ is positive, zero, or negative for a 
spacetime with spherical, flat,
 and hyperbolic spatial geometry, respectively.
 To achieve this 
we first design the fluid flow so that it models a 
transformation from
  Minkowski 
to conformal 
coordinates  $(t,r,\vartheta,\varphi)$ using the prescription of Ibison \cite{ibison},
which we review in appendix \ref{coordinate}. 
In the next section we describe the kinematics of the fluid flow
suitable for modeling an FRW spacetime.

\subsection{Kinematics of the flow}
\label{kinematics}
Applying the transformation
(\ref{eq005})--(\ref{eq006}),
we arrive at 
the line element 
\begin{equation}
ds^2=a^2(t,r)\left(dt^2-dr^2-\frac{\sin^2(\sqrt{k}r)}{k}d\Omega^2\right),
\label{eq1004}
\end{equation}
where $a^2(t,r)$  is given by 
(\ref{eq007}). The particular choice 
$\gamma=\delta=1$ in (\ref{eq007}) gives
\begin{equation}
a^2(t,r) = \left(\cos(\sqrt{k}r)+
\cos(\sqrt{k}t)\right)^{-2}.
\label{eq3004}
\end{equation}
The new temporal and radial coordinates $t$ and $r$  and the spatial curvature radius
$|k|^{-1/2}$ are measured in units
of $m^{-1}$, where $m$ is  
a yet unspecified  mass scale.
Hence, in units of $m^2$ the curvature equals  $k = 1$, 0, or $-1$.

Since the conformal factor $a^2$ is a function of both $t$ and $r$, the line element 
(\ref{eq1004})
is generally  not FRW.
However, for $k=-1$, the expression for $a$ may be simplified 
by taking the limit $\delta \rightarrow 0$ in 
(\ref{eq007}),
in which case we obtain 
$a$ as a function of $t$ only,
\begin{equation}
a(t) = e^{\epsilon t} ,
\label{eq3005}
\end{equation}
where $\epsilon$ may take the value $+1$ or $-1$, corresponding 
to an expanding or  collapsing universe, respectively.

The fluid which provides the desired map 
(\ref{eq005})--(\ref{eq006}) from $(T,R)$ 
to $(t,r)$ coordinates
is required to be at rest in 
$(t,r)$ coordinates. In other words 
the expansion of the fluid is such that the new coordinate frame 
$(t,r)$ is comoving, i.e., 
the four-velocity of the flow in the new coordinate frame is 
\begin{equation}
u_\mu=(a,0,0,0), \hspace{1cm} u^\mu=(1/a,0,0,0).
\label{eq1115}
\end{equation}
Hence,  the inverse transformation of (\ref{eq005})--(\ref{eq006}) applied
to  (\ref{eq1115}) yields 
the velocity components $u_T$ and $u_R$ in  
the original Minkowski frame. 
Expressed in terms of $t$ and $r$ these components 
are given by 
(\ref{eq015}) and (\ref{eq016}).
Specifically, for $\delta = 1$ we have 
\begin{equation}
u_T= \frac{1+ \cos(\sqrt{k}r)\cos(\sqrt{k}t)}{\cos(\sqrt{k}r)+\cos(\sqrt{k}t)},
\label{eq013}
\end{equation}
\begin{equation}
u_R= \frac{\sin(\sqrt{k}r)\sin(\sqrt{k}t)}{\cos(\sqrt{k}r)+\cos(\sqrt{k}t)}.
\label{eq014}
\end{equation}

Again, for $k=-1$ we may take the limit $\delta\rightarrow 0$ and
 the expressions
(\ref{eq015}) and (\ref{eq016}) reduce to
\begin{equation}
u_T=\cosh(\epsilon\, r),
\label{eq017}
\end{equation}
\begin{equation}
u_R=\sinh(\epsilon\, r),
\label{eq018}
\end{equation}
where $\epsilon$ equals $+1$ or $-1$, corresponding 
to an expanding or  collapsing fluid, respectively.
These velocities describe a spherically symmetric Bjorken-type expansion. 
This type of expansion has been recently applied to mimic
an open FRW metric 
 in a relativistic BE system \cite{fagnocchi}
and in a  hadronic fluid \cite{bilic2,bilic3}. 

The above discussed limit $\delta \rightarrow 0$ that removes the $r$-dependence 
of the scale factor in (\ref{eq1004}) does not work for $k=+1$ or $0$.
 However, as we will shortly demonstrate,
with the appropriate choice of the fluid Lagrangian 
we can eliminate $r$-dependence  even for a more general 
expansion of the form (\ref{eq013}) and (\ref{eq014}) applicable to any $k= 1,0,$ or 
$-1$.

Next, we  derive a Lagrangian 
that describes a fluid capable of modeling FRW spacetime 
in terms of the 
relativistic acoustic geometry.
\subsection{Mapping to an analogue FRW spacetime}
%\label{mapping}
%\section{Lagrangian formulation}
\label{lagrangian}

%and the energy-momentum

 For our purpose it proves advantageous to use the field-theoretical description of
 fluid dynamics \cite{garriga}. 
  Consider a Lagrangian ${\cal L}(X,\theta)$
 that depends 
 on a dimensionless scalar field  $\theta$ 
 and on the kinetic energy term 
 \begin{equation}
X = g^{\mu \nu} {\theta}_{, \mu}
\theta_{, \nu}.
\end{equation}
For $X>0$, the energy-momentum tensor  
\begin{equation}
T_{\mu\nu}= 2{\cal L}_X
\theta_{,\mu}\theta_{,\nu}
-{\cal L}g_{\mu\nu} ,
\label{eq508}
\end{equation}
takes the perfect fluid
form 
\begin{equation}
T_{\mu\nu}=(p+\rho) u_{\mu}u_{\nu}-p g_{\mu\nu},
\label{eq019}
\end{equation}
in which 
\begin{equation}
 u_\mu=\frac{\partial_\mu \theta}{\sqrt{X}}.
 \label{eq4012}
\end{equation}
Hence, the field $\theta$ serves as the velocity potential.
 The quantities  
\begin{equation}
p ={\cal L}\, 
\label{eq4003}
\end{equation}
and
\begin{equation}
\rho = 2 X {\cal L}_{X}-{\cal L} 
\label{eq4004}
\end{equation}
are identified as 
 the  pressure and 
energy density of the fluid, respectively,
and the field equation 
\begin{equation}
(2 {\cal L}_X \, g^{\mu\nu} \theta_{,\mu})_{;\nu}
-\partial{\cal L}/\partial\theta
=0
\label{eq400}
\end{equation}
is equivalent to
the continuity equation.
The subscript $X$ in (\ref{eq508})--(\ref{eq400}) 
denotes a partial derivative with respect to $X$.

Small perturbations in the fluid
 propagate in an effective 
curved geometry with 
metric $G_{\mu\nu}$ which may be derived as follows
\cite{babichev,kang}. 
Consider a small perturbation $\chi$ around the 
classical solution  to the field equation (\ref{eq400}).
Replacing
\begin{equation}
\theta(x)\rightarrow \theta(x) +\chi(x) 
 \label{eq2}
\end{equation}
and keeping only the terms quadratic in the derivatives of $\chi$
in the Taylor expansion of the Lagrangian
we obtain the effective action 
\begin{equation}
\delta S =
\frac{1}{2}\int d^4x \sqrt{-G}\,
m^2 G^{\mu\nu}\partial_{\mu}\chi
\partial_{\nu}\chi  ,
\label{eq2007}
\end{equation}
where
\begin{equation}
G^{\mu\nu}=\frac{m^2 c_s}{2{\cal L}_X}
[g^{\mu\nu}-(1-\frac{1}{c_s^2})u^\mu u^\nu] ,
\label{eq2208}
\end{equation}
with $u^\mu=g^{\mu\nu}u_\nu$ and $u_\nu$ as defined in
(\ref{eq4012}). 
The matrix $G^{\mu\nu}$ is the inverse of
the effective metric tensor 
\begin{equation}
G_{\mu\nu}=\frac{{2\cal L}_X}{m^2 c_s}
[g_{\mu\nu}-(1-c_s^2)u_\mu u_\nu]\, ,
\label{eq3008}
\end{equation}
with determinant $G\equiv \det G_{\mu\nu}$.
The mass parameter $m$
in (\ref{eq2007})--(\ref{eq3008})
is introduced to make $G_{\mu\nu}$ dimensionless and 
the quantity $c_s$ is the so-called ``effective'' speed of sound,
defined as
\begin{equation}
c_s^{2}=\frac{{\cal L}_X}{{\cal L}_X+2 X{\cal L}_{XX}}. 
\label{eq2011}
\end{equation}
Hence, the linear perturbations $\chi$ propagate in the effective metric 
(\ref{eq3008})
and the propagation is governed by the equation of motion
\begin{equation}
\frac{1}{\sqrt{-G}}
\partial_{\mu}
\left(
{\sqrt{-G}}\,G^{\mu\nu}
 \partial_{\nu}\chi\right) + \dots
=0.
\label{eq5}
\end{equation}
\\
The basic mechanism that leads to the the effective action of the form (\ref{eq2007}) with 
(\ref{eq2208}) was first noticed by Unruh \cite{unruh} who was also the first 
     to point out that a supersonic flow may cause analog Hawking radiation.

Due to (\ref{eq1115}) and  (\ref{eq4012}), the field $\theta$
 in comoving coordinates $(t,r,\vartheta,\phi)$
is a function of $t$ only
and the kinetic variable $X$ is a function of both $t$ and $r$, 
\begin{equation}
 X=g^{00}\dot{\theta}^2=\frac{\dot{\theta}^2}{a^2} ,
 \label{eq3111}
\end{equation}
where the overdot denotes a derivative with respect to $t$.
The effective metric (\ref{eq3008}) in comoving coordinates takes the form
%\begin{widetext}
\begin{equation}
G_{\mu\nu} = \frac{2{\cal L}_X a^2}{m^2 c_s}
\left(\begin{array}{cccc}
  c_s^2  &          &  &   \\
           & -1  &     &     \\
           &          & -\sin^2 (\sqrt{k}r)/k &   \\
           &          &                  & - \sin^2\vartheta  \sin^2 (\sqrt{k}r)/k  
\end{array} \right).
\label{eq243b}
\end{equation}
%\end{widetext}
A comoving observer receives information transmitted by acoustic perturbations
propagating in the analog spacetime
with the above effective acoustic metric.  

To construct an FRW metric we must get rid of the
$r$ dependence of the conformal factor, i.e., we must  choose a  Lagrangian
with a functional dependence on $X$  such that the
factor $a^2$ in (\ref{eq243b}) is canceled.
Owing to (\ref{eq3111}) we immediately conclude that
${\cal L}_X$ must linearly depend on $X$ and hence
 our Lagrangian must be quadratic in $X$, i.e., we may choose
\begin{equation}
{\cal L} = V(\theta) X^2 ,
\label{eq3113}
\end{equation}
where $V$ is an arbitrary function of $\theta$.
This Lagrangian\footnote{Note that the Lagrangian (\ref{eq3113}) is invariant 
under the conformal transformation of the metric and hence
 it describes a conformal fluid, which may also be seen by 
verifying that the  corresponding energy-momentum tensor is traceless.}
 belongs to a wider class 
of the form ${\cal L}={\cal L}(\tilde{X})$, where $\tilde{X}=f(\theta) X$ 
and $f(\theta)$ 
is an arbitrary positive function of $\theta$. By the field transformation
$\tilde{\theta}_{,\mu}=f^{1/2}\theta_{,\mu}$ the variable $\tilde{X}$ 
becomes the kinetic term of the field $\tilde{\theta}$ and ${\cal L}(\tilde{X})$
takes the form of a purely kinetic $k$-essence
which describes an isentropic fluid \cite{piattella}.

From (\ref{eq3113}) and using (\ref{eq2011}) we  find
\begin{equation}
c_s^2=\frac13 .
\label{eq3114}
\end{equation}
A further restriction on the fluid functions $p$ and $\rho$ is imposed by the field equation  
(\ref{eq400}).  Applying (\ref{eq400}) to (\ref{eq3113}) 
with $\theta=\theta(t)$, we obtain
\begin{equation}
\dot{\theta}^3\left( 4V  \ddot{\theta}
 +\dot{V}\dot{\theta} \right)=0.
\label{eq401}
\end{equation} 
By inspection it may be verified that this equation, besides the trivial solution $\theta={\rm const}$,
 has 
a solution that satisfies
\begin{equation}
V(\theta) \dot{\theta}^4= \kappa m^4 ,
\label{eq402}
\end{equation} 
where $\kappa$ is an arbitrary dimensionless constant.
Eliminating $V$ from (\ref{eq3113}) by (\ref{eq402})
and using (\ref{eq4003}), (\ref{eq4004}), and (\ref{eq3111})
 we find
\begin{equation}
p=\frac{\kappa m^4}{a^4}, \hspace{1cm} 
\rho=\frac{3\kappa m^4}{a^4}.
\label{eq403}
\end{equation} 
Then, the acoustic line element corresponding to the metric (\ref{eq243b}) 
takes the desired form 
\begin{equation}
ds^2=\alpha(\tau)^2\left(d\tau^2-dr^2-\frac{\sin^2(\sqrt{k}r)}{k}d\Omega^2\right),
\label{eq3025}
\end{equation} 
where  a transition to  the conformal time is made
by the replacement
 $\tau =c_s t$.
 With a convenient choice $\kappa = (c_s/4)^2$ 
the cosmological scale factor reads
\begin{equation}
\alpha(\tau)= V(\theta(\tau/c_s))^{1/4}.
\label{eq3026}
\end{equation} 
 In this way, we have shown that the acoustic analog metric takes the conformal form of 
a general FRW spacetime with positive, negative or zero spatial curvature 
depending on the choice of $k$ in the flow velocity  
(\ref{eq013})--(\ref{eq014}).  The evolution of the cosmological scale is 
 determined by the shape of the potential $V$ in the Lagrangian (\ref{eq3113}).

%\subsection{Examples}

It is instructive to study a few simple examples.
\subsubsection*{Static universe}
The simplest possible case is the analog static spacetime
with potential $V=1$, i.e., with the Lagrangian 
\begin{equation}
{\cal L}=X^2
\label{eq4000}
\end{equation}
It is worth noting that an effective Lagrangian of this form has been
recently studied in the context of 
relativistic superfluidity \cite{mannarelli}.
  From (\ref{eq402})  it follows that
 $\theta= \kappa^{1/4} mt$. Hence, if the fluid  described by the Lagrangian 
(\ref{eq4000})
expands according to 
(\ref{eq013})--(\ref{eq014}) (or (\ref{eq017})--(\ref{eq018})
in the $k=-1$ case), a comoving observer perceives 
a static open ($k=-1$) or closed ($k=1$) analog universe. 

\subsubsection*{Analog de Sitter spacetime}
\label{ds}
We next consider  the de Sitter (dS) geometry which admits all three representations
with the corresponding line elements \cite{ibison},
\begin{equation}
ds^2=\left\{ \begin{array}{ll}
\tau^{-2}(d\tau^2-dr^2-r^2d\Omega^2), & \mbox{$k = 0$},\\
\cos^{-2} \tau \left(d\tau^2-dr^2-\sin^2 r d\Omega^2\right), & \mbox{$k = 1$},\\
\sinh^{-2} \tau\left(d\tau^2-dr^2-\sinh^2 r d\Omega^2\right), & \mbox{$k = -1$}.\end{array} \right.\
\label{eq2004}
\end{equation}
Comparing these with (\ref{eq3025}) together with (\ref{eq402}) and  (\ref{eq3026}), we obtain
$V$ and $\dot{\theta}$ as functions of $\tau$ for each $k$.
Replacing $\tau=c_st$ and integrating $\dot{\theta}$ over $t$,
we obtain 
 \begin{equation}
 4\sqrt{c_s}\,\theta(t) = \left\{ \begin{array}{ll}
           c_s^2 t^2 , & \mbox{ $k = 0$},\\  
          2\sin c_s t , & \mbox{$k = 1$},\\
          2\cosh c_s t , & \mbox{$k = -1$}.\end{array} \right.\
\label{eq2005}
\end{equation}
 Plugging these functions back into $V(t)$,
we find 
\begin{equation}
\ V_{\rm dS}(\theta) =\frac{\sqrt{3}}{16\theta^2}  
\label{eq2006}
\end{equation}
for $k=0$ and  
\begin{equation}
\ V_{\rm dS}(\theta) =\frac{3}{(\sqrt{3}-4\theta^2)^2}  
\label{eq2008}
\end{equation}
for both $k=1$ and $k=-1$ with the conditions
$\theta^2 < \sqrt{3}/4$ for $k=1$ and $\theta^2 > \sqrt{3}/4$
for $k=-1$.

As demonstrated by Ibison \cite{ibison}, the transformation (\ref{eq005})--(\ref{eq006})
with $\delta=1$, $\gamma=2$, and $k=\pm 1$
takes the dS line element in the $k=0$ representation
\begin{equation}
ds^2=\frac{1}{T^2}(dT^2-dR^2-R^2d\Omega^2)
\label{eq002}
\end{equation}
to that in the $k=\pm 1$ representations of the type (\ref{eq3025}) with
\begin{equation}
\alpha^2= \frac{k}{\sin^2 (\sqrt{k} \tau)}.
\label{eq003}
\end{equation}
Hence, the inverse coordinate transformation back to the original Minkowski coordinates
takes our acoustic line element (\ref{eq3025}) with (\ref{eq003}) to the conformal spatially 
flat dS line element (\ref{eq002}),
as it should because, by our construction, 
the acoustic metric of the fluid at rest
(i.e., with $k=0$ velocity field)  described by the potential (\ref{eq2006})
is $k=0$ de Sitter in the conformal form. 
\subsubsection*{Analog anti--de Sitter spacetime}
\label{ads}
Anti--de Sitter (AdS) spacetime is expressible only in the $k= -1$ representation
with line element \cite{ibison}
\begin{equation}
ds^2=\frac{1}{\cosh^2 \tau}\left(d\tau^2-dr^2-\frac{\sinh^2 r}{k}d\Omega^2\right).
\label{eq2010}
\end{equation}
Repeating the above procedure, we find 
\begin{equation}
\ V_{\rm AdS}(\theta) =\frac{3}{(\sqrt{3}+4\theta^2)^2} .  
\label{eq3010}
\end{equation}

\section{Bose-Einstein condensate}
\label{bose}
In the context of relativistic acoustic geometry, 
an interesting example of a relativistic fluid is 
the BE condensate which
 has been suggested in a recent paper \cite{fagnocchi}
as a model for an analogue FRW universe.

We first show that, under certain assumptions,
 self-interacting complex scalar field theories
are equivalent to purely kinetic $k$-essence models
(for details see \cite{bilic6}).
We will then analyze the acoustic metric 
and demonstrate that an appropriately chosen $k$-essence Lagrangian  
can simulate  arbitrary
$k=0$ and $k=-1$ FRW model. 
As for $k=1$, 
purely kinetic $k$-essence Lagrangians are not suitable
for the simulation of $k=-1$ FRW models except for a closed static universe,
as in the first example of section \ref{lagrangian}.

%\subsection{Thomas-Fermi correspondence}
%\label{correspondence}
Consider the
Lagrangian
\begin{equation}
{\cal{L}}
 =  g^{\mu \nu} {\Phi^*}_{, \mu}
\Phi_{, \nu}
 - U(|\Phi|^{2}/m^2)
\label{eq1101}
\end{equation}
for  a complex scalar field,
where the potential $U$ is an arbitrary function of 
$|\Phi|^2/m^2$.
The field $\Phi$ satisfies the Klein-Gordon (KG) equation 
\begin{equation}
 (g^{\mu \nu} {\Phi}_{, \nu})_{;\mu}
 + \frac{dU}{d|\Phi|^2} \Phi=0.
\label{eq2103}
\end{equation}
Using the representation
\begin{equation}
\Phi = \frac{\phi}{\sqrt{2}}\exp (-i\theta),
\label{eq1102}
\end{equation}
the Lagrangian (\ref{eq1101}) may be recast into the form
\begin{equation}
{\cal{L}}
% = \frac12 m^2 \phi^2 g^{\mu \nu} \theta_{, \mu}\theta_{, \mu}
= \frac12  \phi^2 g^{\mu \nu} \theta_{, \mu}\theta_{, \mu}
 +\frac12 g^{\mu \nu} \phi_{, \mu}\phi_{, \nu}
 -U(\frac{\phi^2}{2m^2}).
\label{eq2101}
\end{equation}
The KG equation (\ref{eq2103}) 
splits into two equations for the real scalar fields $\theta$ and $\phi$,
\begin{equation}
(Y g^{\mu \nu} \theta_{,\nu})_{;\mu}=0,
\label{eq2206}
\end{equation}
\begin{equation}
%(g^{\mu \nu} \phi_{, \nu})_{;\mu}+ m^2\phi(U_{Y}-X)=0
(g^{\mu \nu} \phi_{, \nu})_{;\mu}+ \phi \left(U_{Y}/m^2-X \right)=0 ,
\label{eq2207}
\end{equation}
where 
we have introduced the abbreviations
\begin{equation}
X = g^{\mu \nu} {\theta}_{, \mu}
\theta_{, \nu}\, ,
 \;\;\;\;\;\;
Y=\frac{\phi^2}{2m^2} \, ,
\label{eq1104}
\end{equation}
and $U_Y$ denotes the derivative $dU/dY$.
A nontrivial classical solution to either equation (\ref{eq2103})
or equations (\ref{eq2206})--(\ref{eq2207}) is called the relativistic 
Bose-Einstein condensate. 
The field configuration 
thus obtained corresponds to a two-component relativistic fluid.

The formalism of acoustic geometry is derived assuming
a perfect irrotational fluid. 
Here, the energy-momentum tensor corresponding to the Lagrangian
(\ref{eq1101}) represents a combination of  two perfect fluids and
generally cannot be put in the form of a single perfect fluid.
However, 
if the spacetime variations of $|\Phi|$ are
small on the scale smaller than $m^{-1}$, i.e.
assuming
$\phi_{, \mu} \ll m \phi$ ,
then the
Thomas-Fermi (TF) approximation \cite{jaksch,parkins}
(or the eikonal  approximation \cite{fagnocchi})
 applies. In this case, if $dU/d|\Phi|^2>0$
 the fluid becomes perfect and 
  the corresponding Lagrangian is equivalent to a Lagrangian
 that depends only on the kinetic term
$X$ \cite{bilic6}. This may be seen as follows. 
 
 The TF approximation amounts to neglecting the kinetic term
$g^{\mu \nu} {\phi}_{, \mu}
\phi_{, \nu}$  in 
(\ref{eq2101}), in which case the Lagrangian becomes
\begin{equation}
%{\cal{L}}_{\rm TF}/m^4= XY -U(Y)\, ,
{\cal{L}}_{\rm TF}=m^2 XY -U(Y) .
 \label{eq1103}
\end{equation}
The field equation (\ref{eq2206}) remains the same,
whereas (\ref{eq2207}) reduces to
\begin{equation}
m^2X- U_Y=0 .
\label{eq1204}
\end{equation}
Now, the energy-momentum tensor corresponding to
the Lagrangian (\ref{eq1103})
takes the perfect fluid
form (\ref{eq019})
with (\ref{eq4012}) and
\begin{equation}
\rho = Y U_Y+ U ,
 \hspace{1cm}   
 p = Y U_Y - U  .
\label{eq1216}
\end{equation}
Obviously, the perfect-fluid description applies
only if  $U_Y >0$ 
which, generally, need not be true for the entire range $0\leq  Y \leq \infty$.
Hence,   the BE fluid in the TF approximation is perfect
for those $Y$ for which $dU/d|\Phi|^2>0$. 

Owing to (\ref{eq1204}) and the obvious relation
\begin{equation}
 {\cal{L}}_X=m^2 Y ,
 \label{eq1106}
\end{equation}
 equation (\ref{eq1103}) together with (\ref{eq1204}) and (\ref{eq1106}) may be 
interpreted  as a Legendre
transformation,
\begin{equation}
{\cal L} (X)=
   m^2XY -U(Y).
  \label{eq1105}
\end{equation}
For a given $U$  the Lagrangian 
${\cal L}(X)$ can be found by solving (\ref{eq1204}) for $Y$
 and plugging the solution into (\ref{eq1105}).
Similarly, if ${\cal L}(X)$ is known, the potential $U$ may be derived 
by solving (\ref{eq1106}) for $X$
 and plugging the solution into (\ref{eq1105}).

From the Lagrangian ${\cal L}(X)$ 
 which now depends only on the kinetic term $X$ 
we find the equation of motion for the field $\theta$,
\begin{equation}
({\cal L}_X g^{\mu \nu} \theta_{, \nu})_{;\mu}=0,
\label{eq1304}
\end{equation}
which is equivalent to (\ref{eq2206}).
However, one should bare in mind that
the field theories
described by the Lagrangians (\ref{eq1103}) and (\ref{eq1105}), respectively,
are equivalent
only at the classical level. 
%since the Lagrangian (\ref{eq1105}) is obtained 
%from (\ref{eq1103}) by eliminating one degree of freedom with 
%help of the equation of motion (\ref{eq1204}).
The energy-momentum tensor constructed
from (\ref{eq1105})
is of the 
form (\ref{eq019}), with the parametric equation
of state
\begin{equation}
\rho = 2X {\cal L}_X -{\cal L},
 \hspace{1cm} p = {\cal L}.
\label{eq1316}
\end{equation}
This equation is just a different parametrization of the 
equation of state (\ref{eq1216}) which
may be easily
verified by using (\ref{eq1204}) to substitute $U_Y$  for  $X$ in (\ref{eq1316}).

In the following we study an expanding fluid in terms of the Lagrangian  
${\cal L}(X)$ as a framework for  an analog model of an FRW universe.
As before, the effective geometry in comoving coordinates is defined by
the  acoustic metric 
(\ref{eq243b}).
  However, the previous procedure of eliminating $a^2$ in the conformal factor
 cannot be applied here\footnote{The only exception is ${\cal L}(X)=X^2$, which describes a static
analog universe as in the first example of section \ref{lagrangian}.}, 
since ${\cal L}(X)$ is now a general function of $X$.
  Nevertheless, in the case of hyperbolic spatial geometry ($k=-1$) 
the general coordinate transformation (\ref{eq005}) and (\ref{eq006}) to comoving coordinates
 may be simplified taking the limit $\delta \rightarrow 0$.
 In this case the transformation   
(\ref{eq317})--(\ref{eq318}) takes the Minkowski spacetime to the Milne spacetime with
the conformal factor $a=e^{\pm t}$ being a function of $t$ only.
This coordinate transformation corresponds to the Bjorken-type spherical expansion
discussed already in a similar context \cite{fagnocchi,bilic2,bilic3}.
In this case, $X$ is a function of $t$ only and likewise ${\cal L}_X $ and
$c_s$.
The effective acoustic line element in comoving coordinates,
\begin{equation}
ds^2= \frac{2{\cal L}_X }{m^2 c_s}e^{2\epsilon t}\left(c_s^2dt^2-dr^2-\sinh^2 r d\Omega^2\right)
\label{eq1317}
\end{equation}
can be  immediately cast into
the conformal FRW form for any 
${\cal L}={\cal L}(X)$
by the transformation to the conformal time,
\begin{equation}
\tau=\int dt c_s ,
\label{eq1320}
\end{equation}
where $c_s$ is defined in (\ref{eq2011}).
Therefore, by choosing an appropriate interaction potential $U$ (or equivalently, 
a $k$-essence Lagrangian ${\cal L}(X)$) 
 one can, in principle, mimic the open hyperbolic 
FRW universe of arbitrary kind. 
This procedure, albeit straightforward, is slightly more involved than 
the one explained in section \ref{spacetimes} because the speed of sound is 
generally not constant.

Given a $k$-essence Lagrangian ${\cal L}(X)$ it is relatively easy to find
the corresponding analog FRW spacetime. 
First, from the field equation (\ref{eq1304}) we find the relation
\begin{equation}
{\cal L}_X =m^3e^{-3\epsilon t} X^{-1/2},
\label{eq1318}
\end{equation}
from which we can express $X$, ${\cal L}(X)$, and $c_s$ in terms of $t$.
These in turn yield a functional dependence on $t$ of the conformal factor
\begin{equation}
\alpha^2\equiv\frac{2{\cal L}_X }{m^2 c_s}e^{2\epsilon t}
\label{eq1325}
\end{equation}
 in the line element (\ref{eq1317}).
Then, using (\ref{eq2011}) and (\ref{eq1320}),
we find $t$ as a function of the conformal time $\tau$, which in turn yields 
$\alpha=\alpha(\tau)$.

The reverse procedure is slightly more involved.
Suppose we want to model a particular hyperbolic FRW spacetime
of the form 
\begin{equation}
ds^2= \alpha(\tau)^2\left(d\tau^2-dr^2-\sinh^2 r d\Omega^2\right).
\label{eq1321}
\end{equation}
We need to find  the Lagrangian $\cal{L}$
such that the conformal factor of the corresponding acoustic line element (\ref{eq1317})
is equal to $\alpha(\tau)$, i.e, 
we require (\ref{eq1325}).
First, from 
(\ref{eq1318})
and using (\ref{eq2011}), we find
\begin{equation}
c_s^2 =-\frac{\epsilon}{6 X}\frac{dX}{dt}.
\label{eq1319}
\end{equation}
From this, with (\ref{eq1325}) and  $d\tau= c_s dt$,
we obtain two coupled differential equations, 
\begin{equation}
\frac{dt}{d\tau}=\left(\frac{X}{m}\right)^{1/2}e^{\epsilon t} \alpha(\tau)^2,
\label{eq1322}
\end{equation} 
\begin{equation}
\frac{1}{X}\frac{dX}{d\tau} =-6\epsilon\left(\frac{dt}{d\tau}\right)^{-1},
\label{eq1323}
\end{equation} 
the solution to which yields $t$ as a function of $X$ in a parametric form.
This function can, in principle, be made explicit by eliminating the parameter $\tau$.
Plugging $t(X)$ into (\ref{eq1318}), we obtain ${\cal L}_X$ as a function of $X$ from which we find
 ${\cal L}(X)= \int dX {\cal L}_X(X)$.

We consider two examples, one for each procedure discussed above.

\subsubsection*{Scalar Born-Infeld theory}
As an easily tractable example, consider the BE interaction potential
\begin{equation}
U= m^4( Y^2+Y^{-2}),
\label{eq1326}
\end{equation}
 which, in the Thomas-Fermi limit, is
equivalent to the 
scalar Born-Infeld theory with the $k$-essence type Lagrangian
\cite{bilic6,bilic5}
\begin{equation}
{\cal L}= -m^4 \sqrt{1-X/m^2}.
\label{eq1327}
\end{equation}
This theory is based on the dynamics of a $d$-brane in the $d+1$-dimensional bulk
\cite{jackiw,bilic7} and 
is of particular interest in cosmology  because of  its 
potential for unifying dark energy and dark matter
in a single entity called the Chaplygin gas \cite{bilic5,kamenshchik,fabris,bento}.
Using the relation (\ref{eq1318}), the speed of sound  can 
 be expressed as
\begin{equation}
c_s= (1-X/m^2)^{1/2}=\left( 1+e^{-6\epsilon t} \right)^{-1/2}.
\label{eq1328}
\end{equation}
From this we find the relation between $t$ and the conformal time $\tau$
\begin{equation}
e^{3\epsilon t}=\sinh 3\epsilon \tau ,
\label{eq13298}
\end{equation}
which may be used to express $X$, ${\cal L}(X)$, and $c_s$ in terms of $\tau$.
Then, using (\ref{eq1325}), one finds
\begin{equation}
\alpha(\tau)= (\sinh 3\epsilon \tau)^{-2/3}\cosh 3\epsilon \tau .
\label{eq13299}
\end{equation}

%\subsection{Examples}
%\label{examples1}
\subsubsection*{Exponential expansion}
As an example of the inverse problem that can be solved analytically, consider the scale factor
in the form of an exponential function,
%\end{equation} 
\begin{equation}
\alpha(\tau)= e^{\beta \tau}.
\label{eq1324}
\end{equation} 
Equations (\ref{eq1322}) and (\ref{eq1323}) can be solved by the ansatz
\begin{equation}
\left(\frac{X}{m}\right)^{1/2}e^{\epsilon t} \alpha(\tau)^2={\rm const}\equiv 1/c_s . 
\label{eq2324}
\end{equation} 
One easily finds that the Lagrangian is of the form
\begin{equation}
{\cal L}= m^{4-2\eta} X^\eta ,
\label{eq2325}
\end{equation} 
where the power $\eta$ satisfies
\begin{equation}
(\eta-2)^2 - \beta^2 (2\eta -1)=0 ,
\label{eq2326}
\end{equation} 
and the speed of sound is given by
\begin{equation}
c^2=(2\eta -1)^{-1}. 
\label{eq2327}
\end{equation}
Obviously, physical solutions must satisfy $\eta\geq 1$.
The Milne universe ($\beta=\pm 1$) is obtained with $\eta=1$ and $\eta=5$
and the static hyperbolic universe ($\beta=0$)  with $\eta=2$.
The Lagrangian (\ref{eq2325}) with  $\eta=2$  belongs to the class discussed in
section \ref{spacetimes} and may also be used to mimic a static closed universe ($k=1$)
with the help of the expansion (\ref{eq013})--(\ref{eq014}).

\section{Hydrodynamic picture}
\label{hydrodynamic}
Now we move from the field-theoretical description of fluid dynamics
to the standard relativistic hydrodynamic formalism to demonstrate 
less formally and perhaps more convincingly that an
 expanding fluid in Minkowski spacetime can be  mapped into  
  an analog FRW spacetime of any spatial curvature.

First we note that if we identify the particle number density as
\begin{equation}
 n=2\sqrt{X}{\cal L}_X.
 \label{eq2112}
\end{equation}
and the specific enthalpy as
\begin{equation}
 w=\frac{p+\rho}{n}=\sqrt{X}
 \label{eq2111}
\end{equation}
 the effective metric (\ref{eq243b}) takes the form
 \begin{equation}
G_{\mu\nu}=\frac{n}{m^2w c_s}
[g_{\mu\nu}-(1-c_s^2)u_\mu u_\nu]
\label{eq2009}
\end{equation}
which, up to the normalization factor $1/m^2$, equals the acoustic metric derived in Ref.\ \cite{bilic} 
for a perfect
 fluid  potential flow with velocity given by
\begin{equation}
 w u_\mu=\partial_\mu \theta 
 \label{eq2012}
\end{equation} 
and the adiabatic speed of sound $c_s$ defined as
\begin{equation}
 c_s^2=\left.\frac{dp}{d\rho} \right|_{s/n} .
 \label{eq2113}
\end{equation}
It may be shown \cite{bilic4} that
this definition of the speed of sound 
coincides with the definition (\ref{eq2011})
of  the effective speed of sound.

As before, the velocity potential  $\theta$
in comoving coordinates $(t, r)$  is a function of $t$ only and hence 
\begin{equation}
 w u_0=w a = \dot{\theta} .
 \label{eq2013}
\end{equation}
Thus, the acoustic line element in comoving coordinates reads
\begin{equation}
ds^2=\frac{na^3}{m^2 c_s \dot{\theta} }\left(c_s^2dt^2-dr^2-\frac{\sin^2(\sqrt{k}r)}{k}d\Omega^2\right).
\label{eq025}
\end{equation}
Our aim is to make the above line element FRW, i.e., 
to design a fluid for which the conformal factor  
and the speed of sound are functions of $t$
only.
For this purpose, we prove that
 the speed of sound $c_s$ and the conformal factor 
in (\ref{eq025}) 
will not depend on $r$
if and only if the particle number density is a function of $t$ and $r$ of the form
\begin{equation}
 n=\frac{m^3}{a^3}f(t) ,
 \label{eq2014}
\end{equation}
where $f(t)$  is an arbitrary dimensionless function of $t$.

First, suppose  $c_s$ and the conformal factor are functions of  $t$ only.
Then,  it  follows immediately that $n$ is of the form (\ref{eq2014}) with $f=c_s\dot{\theta}/m$.
To prove the reverse, suppose Eq.  (\ref{eq2014})  holds.
Then clearly, the conformal factor will not depend on $r$ if $c_s$ does not depend on $r$.
Hence, it is sufficient to 
show that $c_s$ is independent of $r$.
First we note that
\begin{equation}
 p+\rho=wn =\frac{m^3}{a^4}f(t)\dot{\theta},
  \label{eq2015}
\end{equation}
as a consequence of (\ref{eq2013}) and (\ref{eq2014}). 
This in turn implies that the pressure and density are generally of the form
\begin{equation}
 p=\frac{f_p(t)}{a^4}+\chi(t,a), \hspace{1cm}
 \rho=\frac{f_\rho(t)}{a^4}-\chi(t,a) ,
  \label{eq3015}
\end{equation}
where $\chi$ is yet unknown function of $t$ and $a$, and $f_p$ and $f_\rho$ are functions of $t$
such that
\begin{equation}
 f_p+f_\rho=m^3f(t)\dot{\theta} .
  \label{eq3016}
\end{equation}
The function $\chi$ is not arbitrary since $p$ and $\rho$ must satisfy the continuity
equation
\begin{equation}
u^\mu \rho_{,\mu}+(p+\rho){u^\mu}_{;\mu}=0,
\label{eq1003}
\end{equation}
which follows from the energy-momentum conservation.
We now demonstrate that $\chi$ is of the form
$\chi= g(t)/a^4$  where $g(t)$ is a function that does not depend on $r$.
Using (\ref{eq1115}) and (\ref{eq3015}),
the continuity equation (\ref{eq1003}) may be written as 
\begin{equation}
 \dot{\chi} = \frac{\dot{f}_\rho}{a^4}+(3 f_p-f_\rho)\frac{\dot{a}}{a^5}.
  \label{eq3017}
\end{equation}
Comparing this with the general expression for the time derivative of $\chi$,
\begin{equation}
 \dot{\chi} = \frac{\partial\chi}{\partial t}+\frac{\partial\chi}{\partial a}\dot{a},
  \label{eq3019}
\end{equation}
we have
\begin{equation}
 \frac{\partial\chi}{\partial t}=\frac{\dot{f}_\rho}{a^4} ,
\hspace{1cm}
\frac{\partial\chi}{\partial a} = \frac{3 f_p-f_\rho}{a^5}.
  \label{eq3020}
\end{equation}
These two equations have a unique solution,
\begin{equation}
\chi = - \frac{3 f_p-f_\rho}{4a^4},
  \label{eq3021}
\end{equation}
and hence $\chi$ is indeed of the form $g(t)/a^4$.
Furthermore,
from (\ref{eq3015}) and (\ref{eq3021}) it follows that 
$c_s^2=p/\rho =1/3$, and hence $c_s$ is independent of $r$, which was to be shown. 
%\hfill {\Large $\Box$}

Without loss  of generality, we may set $\chi\equiv 0$  in (\ref{eq3015}),
in which case equation (\ref{eq3020}) implies
\begin{equation}
3 f_p-f_\rho=0, \hspace{1cm} \dot{f}_\rho=0,
  \label{eq3022}
\end{equation}
yielding  
the previously obtained  expressions for the pressure
and density (\ref{eq403}),
where $\kappa$ is an arbitrary positive dimensionless constant.
Equations (\ref{eq403}) provide a relation between the functions $f(t)$ and $\dot{\theta}$. 
From 
(\ref{eq403}) and  (\ref{eq2015}) we find $\dot{\theta} =4 \kappa m/f(t)$.
Our fluid is then specified by (\ref{eq2014}) and (\ref{eq403}).
Furthermore, if we identify  
$f(t)=4\kappa^{3/4} V^{1/4}$ 
and choose $\kappa=(c_s/4)^2$
as before,  the acoustic line element (\ref{eq025}) 
takes 
precisely the same conformal FRW form  
(\ref{eq3025}) with (\ref{eq3026}).
The evolution of the cosmological scale is 
 determined by the function $f(t)$ that appears in the
expression (\ref{eq2014}) for the particle number density.

\section{Analog horizons}
\label{analog}
In this section we derive the expressions for the Hubble and the apparent horizons
for a general analog FRW spacetime such as that in sections \ref{spacetimes} and \ref{hydrodynamic}. 
The FRW line element considered here is assumed to be in the  conformal form 
(\ref{eq3025}), with the acoustic conformal time $\tau=c_s t$.
 The proper distance $d_{\rm p}$ and the comoving spatial distance $r$ are related by $d_{\rm p}=\alpha r$
 as usual, whereas
the analog Hubble expansion rate in conformal coordinates is given by
\begin{equation}
{\cal H}=\frac{\dot{\alpha}}{\alpha^2},
\label{eq238}
\end{equation}
where the overdot denotes a partial derivative with respect to $\tau$. 
We define 
the {\em analog Hubble horizon}  as a two-dimensional spherical surface at which the 
magnitude of the analog recession velocity
\begin{equation}
v_{\rm rec}\equiv {\cal H}d_{\rm p}  = r\frac{\dot{\alpha}}{\alpha}
\label{eq038}
\end{equation}
equals the maximally allowed velocity of sound, $c_s=1$. Hence, 
the condition
\begin{equation}
r =\frac{\alpha}{\left|\dot{\alpha}\right|}
\label{eq321}
\end{equation}
defines the location of the analog Hubble horizon.

\begin{figure}[t]
\begin{center}
\includegraphics[width=0.7\textwidth,trim= 0cm 0cm 0cm 0cm]{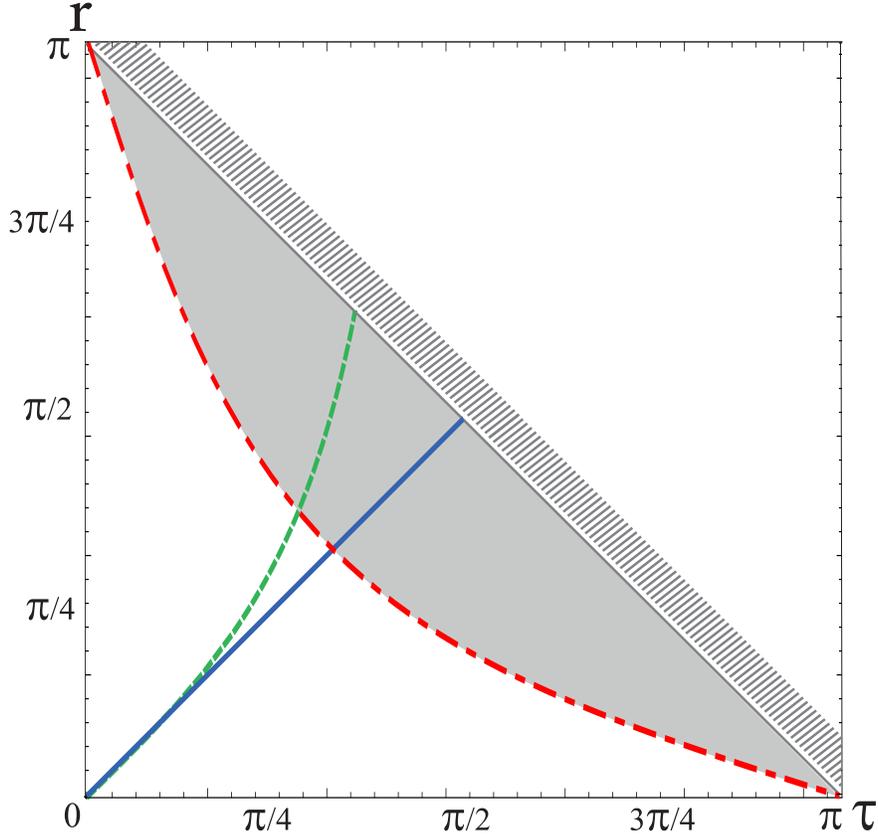}
\caption{Spacetime diagram of the analog $k=1$ dS spacetime
 in $(\tau ,r)$ coordinates. 
The evolution of the apparent, Hubble, and naive horizons
are represented by the full,
dashed, and dash-dotted lines,
respectively. The shaded area represents the region in which
$v\geq c_s$. The diagonal ``wall" marks the boundary of the physical region.}
 \label{fig1}
\end{center}
\end{figure}

Next we define the {\em analog apparent horizon} as a boundary of the 
analog trapped region \cite{bilic3}.
More precisely, the apparent horizon is defined as  
a two-dimensional surface on which one of the null 
expansions vanishes \cite{hayward}.
In the case of spherical symmetry one  can use a more
practical definition:  the apparent horizon is a two-dimensional surface $H$
such that  the vector $n_\mu$, normal to the surface of spherical symmetry
is null on $H$.  In our case it means that $n_\mu$ is null with respect
to the acoustic metric. More explicitly,
the acoustic metric magnitude of the vector 
\begin{equation}
n_\mu =\partial_\mu(\alpha \sin(\sqrt{k}r)/\sqrt{k})
\label{eq319}
\end{equation}
 should be zero on $H$ , i.e.,
\begin{equation}
G^{\mu\nu} n_\mu n_\nu|_H =0.
\label{eq320}
\end{equation}
From  this  one finds the condition for the apparent horizon 
\begin{equation}
\frac{\dot{\alpha}}{\alpha}\pm \sqrt{k}\cot (\sqrt{k}r) = 0,
\label{eq117}
\end{equation}
where $k=1,-1$. 
Any solution to Eq.\ (\ref{eq117}) gives the location of 
the analog apparent horizon $r_H$.

Finally, we define the {\em naive horizon} as a  two-dimensional 
 surface  on which the radial  velocity $v\equiv u_R/u_T$ equals the velocity of sound $c_s$.
 In the case of a stationary spherically symmetric flow, the apparent and naive horizons coincide
 with the analog event horizon \cite{visser}.

\begin{figure}[t]
\begin{center}
\includegraphics[width=0.7\textwidth,trim= 0cm 0cm 0cm 0cm]{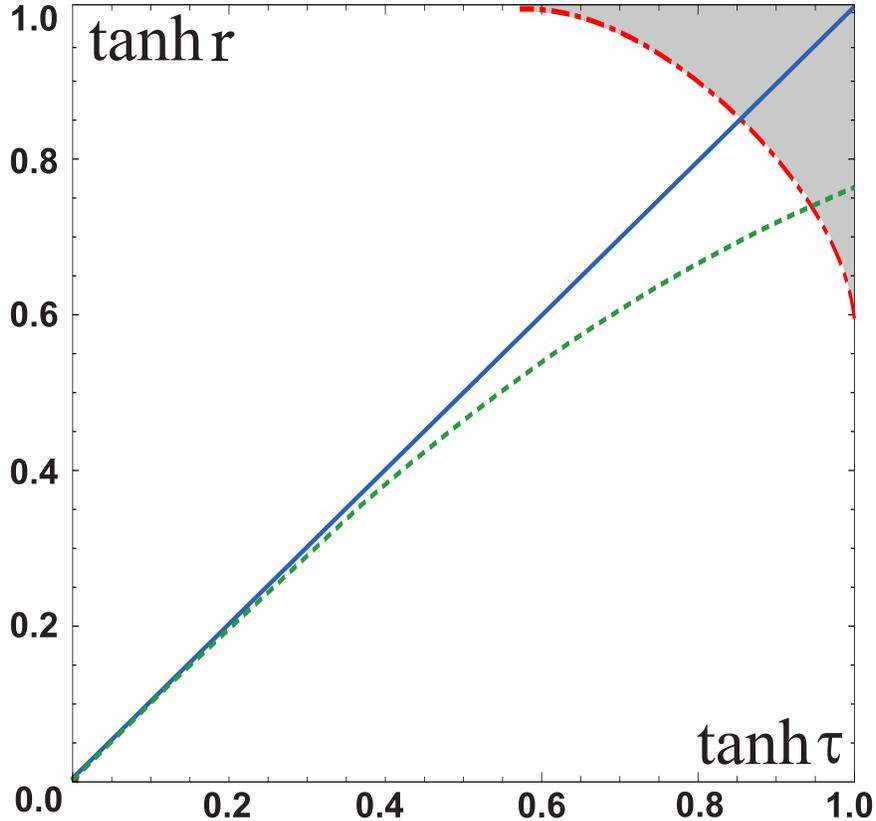}
\caption{Spacetime diagram of the $k=-1$ analog dS spacetime.
The evolution of the apparent, Hubble, and naive horizons
are represented by the full,
dashed, and dash-dotted lines,
respectively.}
 \label{fig2}
\end{center}
\end{figure}

As a simple example, we consider the analog de Sitter spacetime with positive or negative 
spatial curvature. 
We have shown in section \ref{spacetimes} how to mimic  
the $k=1$ and $k=-1$ representations of the dS spacetime: 
 the fluid described by the Lagrangian
(\ref{eq3113}) with the potential (\ref{eq2008}) should be forced to expand according to 
 (\ref{eq013})--(\ref{eq014}) where
 $k$ determines whether the analog spacetime has 
a positive or negative curvature.
In Figs. \ref{fig1} and \ref{fig2} we plot 
the spacetime diagrams in conformal coordinates $(\tau,r)$ 
for the analog dS universe
with positive and negative
spatial curvature, respectively. 
The diagrams depict the evolution of the apparent (full line), analog (dashed line), 
and naive Hubble (dash-dotted line) horizons.
The shaded area in both figures represents the region of supersonic flow, i.e.,
the region in which
the fluid three-velocity $v$ exceeds the speed of sound 
$c_s$. 
The diagonal line $t+r=\pi$ in  Fig. \ref{fig1} 
corresponds to the future infinity, i.e., the point
 $R=T=\infty$ in Minkowski coordinates.

\section{Discussion and conclusions}
\label{discussion}
We have demonstrated that it is possible, at least in
principle, to model an FRW spacetime with arbitrary spatial curvature
using a relativistic fluid and the associated 
effective acoustic geometry.
To obtain the desired geometry one needs to
manipulate the fluid equation of state in addition to
an appropriately designed fluid expansion. 
 We have shown in section \ref{bose} that an open FRW spacetime with zero or negative spatial
curvature can be modeled 
using isentropic perfect fluids such as a Bose-Einstein condensate in the Thomas-Fermi
limit. 
The modeling of a closed FRW
spacetime with positive spatial curvature
is possible if the condition of isentropy is relaxed.
This is achieved by a Lagrangian with a quadratic kinetic energy term
multiplied by a potential $V,$ the choice of which determines 
 the evolution of the analog cosmological scale.
The acoustic analog metric takes the conformal form of 
a general FRW spacetime with positive, negative, or zero spatial curvature 
depending on the choice of the sign parameter $k$ in the expression 
(\ref{eq013})--(\ref{eq014}) for the flow velocity.

It is conceivable that the analysis presented here is  of considerable theoretical 
 interest for emergent gravity models, 
in particular for those in which the emergent gravity is based on fluid flow  
(for a comprehensive review and extensive list of references, see Ref.\ \cite{barcelo2}). 
It is worth mentioning a few  examples in which there is an obvious connection to our analysis.  
The first one concerns already mentioned emergence of scalar gravity.
As was shown in Ref.\ \cite{novello2}, in the framework of the fluid-field correspondence 
one can go beyond analog gravity by introducing a scalar field Lagrangian that describes 
the dynamics of a scalar field as an interaction of the field and its associated 
effective metric given by (\ref{eq3008}). 
This interaction may be interpreted as a gravitational influence 
on the field by its own effective metric 
\cite{novello,novello2}.
 Another example is the model of Janik and Peschanski \cite{janik}
 in which perfect fluid hydrodynamics emerges 
as a consequence of the AdS/CFT correspondence.
In their approach 
the AdS/CFT correspondence relates 
a perfect conformal fluid on the boundary  to an asymptotically AdS$_5$ bulk.
The link to our study is twofold: first, the fluid 
described by the Lagrangian (\ref{eq3113}) 
is conformal and second, the longitudinal Bjorken expansion of Ref.\ \cite{janik} 
is of the same form 
as our spherical expansion (\ref{eq017}) and (\ref{eq018}) obtained 
from the more general expression in the limit when the parameter $\delta$
approaches zero. 
Assuming a true AdS/CFT duality, the boundary conformal fluid in the model of Ref.\ \cite{janik}
 may also be regarded as primary and  the bulk as emergent \cite{carlip}.

In our study so far we have discussed  simple toy models with no
realistic fluid in mind.
To the best of our knowledge, the only realistic experimental set up for a relativistic-fluid laboratory
is provided by high-energy colliders.
In our previous papers \cite{bilic2,bilic3} we suggested a relativistic collision model
based on a spherical Bjorken-type expansion
to mimic an open FRW universe.
To model a closed universe, a different type of expansion is needed with the $k=1$ velocity field
(\ref{eq013})--(\ref{eq014}). Presently we do not have a concrete proposal for
how to do this in high-energy collisions. Maybe, with the advance of accelerator
technology, one day it will be possible, e.g., by choosing appropriate heavy
ions and specially designed beam geometry to obtain
the desired equation of state and expansion flow of the fluid.
To confront the model with experiment and test the properties of the analog spacetime
it would be of interest to investigate  
the effects of  the cosmological particle production in our relativistic model 
in the same way as it was done in the nonrelativistic analog models of expanding universes
\cite{barcelo1,weinfurtner}.
Besides, the presence of a trapped region and the apparent horizon can, in principle, be detected
by looking for a signal of the analog Hawking radiation in the
particle distribution spectra.

\subsection*{Acknowledgments}
This work was supported by the Ministry of Science,
Education and Sport
of the Republic of Croatia under Contract No. 098-0982930-2864
and 
partially supported by the ICTP-SEENET-MTP grant PRJ-09 ``Strings and Cosmology" 
in the frame of the SEENET-MTP Network.
N.\ B.\ thanks CNPq, Brazil, for partial support and the University of Juiz de Fora where a part of 
this work has been completed.

\appendix

\numberwithin{equation}{section}

\section{Transformations to conformal coordinates} 
\label{coordinate}
Here we describe the procedure \cite{ibison} 
for a transformation from Minkowski spherical coordinates
to conformal spatially closed or hyperbolic coordinates.
We start from
the background spacetime of the form
\begin{equation}
ds^2=dT^2-dR^2-R^2d\Omega^2,
\label{eq001}
\end{equation}
and apply the following transformation:
\begin{equation}
T(t,r)=T_0+\frac{\gamma}{2\delta^2\sqrt{-k}}
\left[f(t+r) +f(t-r)\right],
\label{eq005}
\end{equation}
\begin{equation}
R(t,r)=R_0 + \frac{\gamma}{2\delta^2\sqrt{-k}}
\left[f(t+r) -f(t-r)\right],
\label{eq006}
\end{equation}
where
\begin{equation}
f(x)= \tanh \left(\frac{\sqrt{-k}}{2} x+ \log\delta\right),
\label{eq106}
\end{equation}
$T_0$, $R_0$, $\gamma$, and $\delta$ are constants, and $k = 1,0,-1$ 
for the spherical, flat, or hyperbolic spatial geometry, respectively. 
Without loss of generality the offsets $T_0$, $R_0$, to the origins of
$T$ and $R$ may be set to zero with the implicit assumption
that in any result the coordinates $T$ and $R$ can be linearly rescaled. 
This transformation takes the line element (\ref{eq001}) to the conformal form
\begin{equation}
ds^2=a^2(t,r)\left(dt^2-dr^2-\frac{\sin^2(\sqrt{k}r)}{k}d\Omega^2\right),
\label{eq004}
\end{equation}
where 
\begin{equation}
a^2(t,r)=\frac{\gamma^2}{\delta^4\left[\cosh(\sqrt{-k}r)+
\cosh(\sqrt{-k}t + 2\log\delta)\right]^2},
\label{eq007}
\end{equation}
or equivalently, in terms of $T$ and $R$,
\begin{equation}
\ a^2(T,R) =\left[\frac{\gamma}{2\delta^2}-
\frac{\delta^2}{2\gamma}k(T^2-R^2)\right]^2 +k T^2.
\label{eq107}    
\end{equation}
These expressions can be simplified by a convenient choice of 
the constants $\gamma$ and $\delta$.
Specifically, for $\gamma=2$ and $\delta=1$ we obtain
\begin{equation}
\ a^2(t,r) = \left\{ \begin{array}{ll}
         1, & \mbox{$k = 0$};\\
         4/(\cos r + \cos t)^2 , & \mbox{ $k = +1$};\\
         4/(\cosh r + \cosh t)^2, & \mbox{$k = -1$},\end{array} \right.\
\label{eq34}
\end{equation}
\begin{equation}
\ a^2(T,R) = \left\{ \begin{array}{ll}
         1, & \mbox{$k = 0$};\\
         \left[1-(T^2-R^2)/4\right]^2 + T^2 , & \mbox{$k = +1$};\\
         \left[1+(T^2-R^2)/4\right]^2 - T^2 , & \mbox{$k = -1$}.\end{array} \right.\
\end{equation}

%\subsection{Coordinate transformation (2)}

A particularly simple transformation is obtained in
the limit $\delta\rightarrow 0$ in which case one must  
choose a nonzero offset $T_0$ in (\ref{eq005}) to obtain a finite result.
In this limit  
the transformation (\ref{eq005})--(\ref{eq106}) with $T_0=\gamma/(\delta^2\sqrt{-k})$ 
takes the form
\begin{equation}
T=2\gamma\frac{e^{\sqrt{-k}t}}{\sqrt{-k}}\cosh(\sqrt{-k}r),
\label{eq317}
\end{equation}
\begin{equation}
R=2\gamma\frac{e^{\sqrt{-k}t}}{\sqrt{-k}}\sinh(\sqrt{-k}r),
\label{eq318}
\end{equation}
which, obviously, makes sense only for 
 $k = -1$. Then we obtain 
 the conformal representation of the Milne universe,
 \begin{equation}
ds^2=e^{2\epsilon t}\left(dt^2-dr^2-\sinh^2 r d\Omega^2\right)
\label{eq009}
\end{equation}
where $\epsilon$ may take the value $+1$ or $-1$, corresponding 
to an expanding or  collapsing universe, respectively.
This is the only possible nontrivial map from the Minkowski spacetime 
to an FRW spacetime with negative spatial curvature.
A direct map to an FRW space with positive spatial curvature is not
possible with the coordinate transformation (\ref{eq005}).

%\subsection{Transformation of the 4-velocity}
Consider next a spherically expanding (or contracting) fluid such that 
 the $(t,r)$ coordinate frame is comoving, i.e., 
such that the flow velocity of the expanding fluid in that frame is 
\begin{equation}
u_\mu=(a,0,0,0).
\label{eq115}
\end{equation}
Then, the components of the velocity in the $(T,R)$ coordinate frame are
\begin{equation}
u_T = g_{00}^{-1/2} T_{,t} , 
\end{equation}

\begin{equation}
u_R=g_{00}^{-1/2}R_{,t}.
\end{equation}
Using this and (\ref{eq005})--(\ref{eq106}), we find the velocity components expressed in terms of $t$ and $r$, 
%\begin{equation}
%u^{\mu}=\frac{\partial x^{\mu}}{\partial x^{\nu}}u^{\nu},
%\label{eq112}
%\end{equation}
\begin{equation}
 u_T 
= \frac{2\delta^2+\cosh(\sqrt{-k}r)
\left(e^{-\sqrt{-k}t}+e^{\sqrt{-k}t}\delta^4\right)}{2\delta^2
\cosh(-\sqrt{k}r)+e^{-\sqrt{-k}t}+e^{\sqrt{-k}t}\delta^4},  
\label{eq015}
\end{equation}
\begin{equation}
u_R   
= \frac{\sinh(\sqrt{-k}r)
\left(e^{-\sqrt{-k}t}-e^{\sqrt{-k}t}\delta^4\right)}{2\delta^2
\cosh(-\sqrt{k}r)+e^{-\sqrt{-k}t}+e^{\sqrt{-k}t}\delta^4}.
\label{eq016}
\end{equation}
and in terms of $T$ and $R$,
\begin{equation}
u_T= \frac{1}{a (T,R)}\left[\frac{\gamma}{2\delta^2}+
\frac{\delta^2}{2\gamma}k(T^2+R^2)\right] 
\label{eq015b}
\end{equation}
\begin{equation}
u_R= \frac{1}{a (T,R)}\frac{\delta^2}{\gamma}kTR,
\label{eq016b}
\end{equation}
where $a (T,R)$ is given by (\ref{eq107}).

\end{document}